\documentclass[twocolumn, amsmath, aps, fleqn,nofootinbib]{revtex4}

\usepackage{amsmath}
\usepackage{amssymb}
\usepackage{graphicx}
\usepackage{subfigure}
\usepackage{epsfig}
\usepackage{verbatim}
\usepackage{natbib}
\usepackage{eucal}
\usepackage{calligra}
\usepackage{mathrsfs}
\usepackage{enumerate}

\usepackage[usenames,dvipsnames]{color}

\usepackage{amsfonts}
\usepackage{color}
\usepackage[normalem]{ulem}
\usepackage[T1]{fontenc}

\DeclareMathAlphabet{\mathscr}{OT1}{pzc}%
                                 {m}{it}
                                 
\newcommand{\mnras}{MNRAS}
\newcommand{\jcap}{JCAP}
\newcommand{\pasj}{Publications of the Astronomical Society of Japan}
\newcommand{\apjl}{ApJL}

\newcommand{\aap}{A\&A}
\newcommand{\physrep}{PhR}

\newcommand{\be}{\begin{equation}}
\newcommand{\ee}{\end{equation}}
\newcommand{\bes}{\begin{equation*}}
\newcommand{\ees}{\end{equation*}}
\newcommand{\bea}{\begin{eqnarray}}
\newcommand{\eea}{\end{eqnarray}}
\newcommand{\beas}{\begin{eqnarray*}}
\newcommand{\eeas}{\end{eqnarray*}}

\newcommand{\zl}{z_{\rm L}}
\newcommand{\zs}{z_{\rm s}}

\begin{document}

\title[Void Lensing as a Test of Gravity]
{Void Lensing as a Test of Gravity}

\author{Tessa Baker}
\email{tessa.baker@physics.ox.ac.uk}
\affiliation{University of Oxford, Astrophysics, Denys Wilkinson Building, Keble Road, Oxford OX1 3RH, United Kingdom}

\author{Joseph Clampitt}
\email{clampitt@sas.upenn.edu}
\affiliation{Department of Physics and Astronomy, Center for Particle Cosmology, University of Pennsylvania, 209 S. 33rd St., Philadelphia, PA 19104, USA}

\author{Bhuvnesh Jain}
\email{bjain@physics.upenn.edu}
\affiliation{Department of Physics and Astronomy, Center for Particle Cosmology, University of Pennsylvania, 209 S. 33rd St., Philadelphia, PA 19104, USA}

\author{Mark Trodden}
\email{trodden@physics.upenn.edu}
\affiliation{Department of Physics and Astronomy, Center for Particle Cosmology, University of Pennsylvania, 209 S. 33rd St., Philadelphia, PA 19104, USA}

\begin{abstract}
\noindent We propose a consistency test of gravity based on the weak lensing signal of cosmic voids. For a given void profile, as traced by galaxies, the lensing signal can vary in different gravity theories. 
Thus the comparison of the lensing shear profile of such voids with the General Relativistic prediction can test for deviations from GR. For concreteness, we calculate the expected lensing signal in two gravity theories involving scalar fields with derivative couplings. We find that the scalar field has the potential to boost the tangential shear both within and outside the void radius. Reversing the method, one can infer the void central density parameter from the lensing signal, and compare to the value estimated independently using the galaxy tracer profiles of voids. Hence one can check for consistency between the behaviour of light and matter under the assumption of GR.
We use voids traced by Luminous Red Galaxies in SDSS to demonstrate our methodology, finding that the void central density parameter can shift from its GR value by up to 20\% in some galileon gravity models. Although galileon gravity is now disfavoured as a source of cosmic acceleration by other datasets, the methods we demonstrate here can be used to test for more general fifth force effects with upcoming void lensing data.
\end{abstract}

\maketitle

\section{Introduction}

\noindent Gravitational lensing by cosmological voids -- under-dense regions of the universe typically 10--100 Mpc in size -- is  emerging as a promising new tool for studies of dark energy and large-scale structure \cite{Zhao2018}. Since the detection of lensing by stacked voids in SDSS \cite{Krause2013, Melchior2014, Clampitt2015} and  related work on both theory and measurement (\cite{Jennings2013,Nadathur2015a,Hamaus2016,Clampitt2016, Nadathur2017} and references therein), void lensing has been measured in the Dark Energy Survey \cite{Sanchez2017}, with considerable improvements to follow in subsequent data releases. The related phenomenon of trough lensing has  been developed into a cosmological probe as well \cite{Barreira2016, Gruen2016, Friedrich2017}.

In particular, voids have the potential for powerful tests of gravitational `fifth forces', interactions induced by a new fundamental field mediating long-range forces between matter fields. One setting in which such fifth forces are common is in modified theories of gravity. Many extended gravity theories are accompanied by `screening mechanisms' that strongly suppress their observable effects in high-density regions of the universe such as the Solar System and the interior of galaxies \cite{VAINSHTEIN1972393,Khoury2004, Joyce2015}. Though known screening mechanisms (chameleon, Vainshtein, etc.) differ in their precise details, a feature common to all is that high density environments are generally screened. 

In contrast, the low-density nature of voids means that screening mechanisms do not operate inside them, allowing deviations from GR to play at full strength \cite{Falck2015,Falck2018,Cautun2018}. The aim of this paper is to study the ways that a gravitationally-coupled scalar field impacts the lensing profiles of voids, and hence to uncover useful phenomenology for fifth-force constraints with forthcoming galaxy survey data \cite{KiDS, HSC, Euclid, LSST, WFIRST}.

Recent bounds on the gravitational wave propagation speed using a binary neutron star merger~\cite{Renk2017,Baker2017,Zuma2017,Creminelli2017} have strongly constrained modified gravity theories as the mechanism of cosmic acceleration, arguably their original motivation (see also earlier work by \cite{Lombriser2016, Lombriser2017}). However, there remain some classes of theories which do not modify the speed of gravitational waves, and hence evade these constraints. For example:
\begin{enumerate}
\item Theories with non-universal couplings between baryons and dark matter;
\item Scalar-tensor theories in which the background value of the scalar does not evolve at low redshifts. $|\dot{\bar\phi}|\ll 1$;
\item Specialised theories within the generalised families of Horndeski, Beyond Horndeski and DHOST gravity, whose defining Lagrangian functions satisfy certain constraint relations \cite{Gleyzes2015a,Gleyzes2015b,Noui2016,Crisostomi2017,Kase2018};
\item Theories containing an intrinsic mass scale significantly greater than the mass associated to the Hubble scale, $m^2_{\phi}\gg m^2_H$.
\end{enumerate} 
In the lattermost case, the interesting phenomena of the theory are shifted towards astrophysical (sub-cosmological) scales. There is a growing body of work studying the effects of fifth forces on clusters, galaxies and the nonlinear regime of large-scale structure formation~\cite{Burrage:2017qrf,Burrage:2016bwy}. This shift towards astrophysical scales generally has the consequence that such theories cannot successfully accelerate the universe without an additional cosmological constant. 

Fifth forces can arise in many other settings, beyond specific applications to modified gravity and late-time cosmic acceleration. Attempts to address cosmological problems such as the behavior of dark matter on small scales, or baryogenesis, involve new scalar fields with long-range interactions. Likewise, attempts to embed our existing theories in a more fundamental model with extra spatial dimensions can give rise to the similar interactions, either through compactification, or as brane bending modes. Furthermore, as shown in~\cite{Rajantie2014}, during inflation a scalar such as the Higgs field can acquire couplings to the Einstein-Hilbert term at loop order, even if minimally coupled to gravity at tree level. In all such models, the distances/energies/densities at which these deviations from GR are most significant are determined by the mass scales of the underlying fundamental theory. These mass scales could be large enough (and hence the wavelength of the scalar fifth force short enough) for the theory to satisfy cosmological and gravitational wave constraints, yet have non-trivial effects on astrophysical scales.

The relevant physics we are interested in, therefore, is not primarily focused on the question of whether modified gravity can explain cosmic acceleration, but rather is part of the grand program of testing gravity on all possible scales. In particular, our question is whether there may be new light scalar fields propagating on sub-Hubble scales. The inherent complexity of astrophysical systems means that existing constraints here are less developed than cosmological ones; hence voids can provide useful new information in this arena.

When screening is not effective, the new fields that generically accompany modifications to GR will have non-trivial radial profiles across the void. This modifies the weak lensing profile of the void via two channels:
\begin{enumerate}[i)]
\item Because the new field(s) act as an effective extra source of stress-energy, the gravitational potential inside the void is no longer purely the one deduced from baryons and dark matter. At the level of the gravitational field equations, this means that the Poisson equation is modified (our notation for metric potentials is introduced in eq.\ref{lineel}):
\begin{align}
\quad\quad \nabla^2\Phi&=4\pi G \left(\rho_M+\rho_{\rm eff}\right) \ ,
\end{align}
where $\rho_{\rm eff}$ is the effective stress-energy contribution of the new field. 
\item In some models, the new field further acts as an effective source of anisotropic stress. As a result, the lensing potential of the void is no longer equal to the gravitational potential experienced by nearby masses, i.e. \mbox{$\Psi_L = (\Phi+\Psi)/2\neq \Psi$}. 
\end{enumerate}
Clearly these two types of modification are partially degenerate in their resulting effects on the lensing shear profile of the void. To understand their different influences, consider a situation in which the new field $\phi$ is a subdominant component of the energy density, i.e. $\rho_{\rm eff} \ll \rho_M$ (inside the void, at least). Then, the modifications from the first effect above would vanish, while modifications from the second effect would still operate. In this paper we will study two gravity models, corresponding to two different terms in the simple galileon Lagrangian. We will see that one of these models acts via both the above effects, while the other acts only via the first one.

A key input to our following calculations is a model for the radial density profile of a void, which can be obtained from fitting to either simulations or galaxy catalogs. In this paper we will use two density profiles; one is a simple fit from galaxy catalogs, whilst the other has enhanced flexibility of the type identified in simulations in \cite{Hamaus2014}. Comparison with real data will rely on a good understanding of the void-tracer connection; see, for example, work on Halo Occupation Distribution models in \cite{slw2013, slw2014, Nadathur2015b,Pollina2017}. Likewise, a high-accuracy analysis will need to account for the possible impact of deviations from GR on galaxy bias relations; in the present work we neglect these, since they are likely to be subdominant to current bias uncertainties within GR itself \cite{Neyrinck2014,bias2016}.

The structure of this paper is as follows: in \S\ref{sec:models} we introduce the gravity models under study, their background equations and particular features. In \S\ref{section:voidcalc} we present the calculation of the tangential shear profile of a void, the results of which are studied in  \S\ref{section:theory results}. \S\ref{section:discussion} summarizes our results, caveats, and directions for future work.

\section{Gravity Models}
\label{sec:models}

\subsection{Action \& Motivations}
\label{subsec:models}
As a simple framework to study deviations from GR, we restrict ourselves to two terms from the simplest galileon family of gravity theories. Galileon gravity is constructed using a scalar field, $\phi$, which is characterised by its unusual higher-derivative self-couplings~\cite{Nicolis:2008in}. Galileon fields arise in a number of different ways, for example, having elegant geometrical origins as the description of brane fluctuations in the DGP model~\cite{Dvali:2000hr,Luty:2003vm}, and describing the helicity zero component of ghost-free massive gravity~\cite{deRham:2010ik,deRham:2010kj}. Furthermore, they exhibit a rich structure and a complex phenomenology, including Vainshtein screening~\cite{Vainshtein:1972sx} near massive objects~\cite{Nicolis:2004qq,Nicolis:2008in}, and possessing an S-matrix with a number of special properties~\cite{Kampf:2014rka,Cheung:2014dqa,Cachazo:2014xea,Cheung:2015ota,Hinterbichler:2015pqa,Goon:2016ihr}. They also face some theoretical challenges, such as perturbations propagating superluminally around some sources~\cite{Nicolis:2008in,Goon:2010xh}  and the existence of arguments that they have no local, Lorentz invariant UV completions~\cite{Adams:2006sv,Bellazzini2017}, although there are creative attempts to circumvent some of these~\cite{Cheung:2016yqr,Keltner:2015xda}.

Galileons have been of particular interest in recent years as a possible candidate, with a particular choice of parameters, for explaining late-time cosmic acceleration. While the relevant parameter range for that application is now tightly constrained, the models more generally provide a framework for how deviations from GR might exist but remain hidden from local tests of gravity. As we will explain in more detail soon, it is in this spirit that we make use of them in this paper.

We consider the simplest example of galileons -- a single scalar field, $\phi(x)$, which obeys a shift symmetry linear in coordinates:
\begin{equation}
\phi(x) \rightarrow \phi(x)+c+b_\mu x^\mu \ ,
\label{galsym}
\end{equation}
with $c$, $b_{\mu}$ constants.  Any term built out of $\partial_{\mu}\partial_{\nu}\phi$, and its derivatives, will be strictly invariant under eq.\eqref{galsym}.  However, there also exist special operators with {fewer} than two derivatives per $\phi$, which are not strictly invariant, but rather are invariant up to a total derivative.  There are three main examples of these terms. These are the cubic, quartic and quintic models, all of which contain higher derivatives, and which involve three, four and five copies of the field, respectively. In this paper we will work with only the cubic and quartic galileons --  these terms are sufficient to demonstrate the two `channels' for void lensing that we described in the introduction. We stress here that our main interest is the effect of non-GR physics on void lensing observables, and not the detailed specifics of galileon gravity.

The general galileon action up to order 4, but not containing terms built out of $\partial_{\mu}\partial_{\nu}\phi$, is:
\begin{align}
\label{eq:action}
 S =& \int {\rm d}^4x\sqrt{-g} \left[ \frac{R}{16\pi G} - \frac{1}{2}\sum_{i=1}^5c_i\mathcal{L}_i - \mathcal{L}_m\right] \ ,
 \end{align}
 with
 \begin{align}
\mathcal{L}_1 =& M^3\phi \ , \nonumber \\
\mathcal{L}_2 =& \nabla_\lambda\phi\nabla^\lambda\phi  \ , \nonumber \\
\mathcal{L}_3 =& \frac{2}{M^3}\square\phi\nabla_\lambda\phi\nabla^\lambda\phi \ , \nonumber \\
\mathcal{L}_4 =& \frac{\nabla_\lambda\phi\nabla^\lambda\phi}{M^6}\times\nonumber\\
&\Big[ 2(\square\phi)^2 - 2(\nabla_\mu\nabla_\nu\phi)(\nabla^\mu\nabla^\nu\phi) -R\nabla_\mu\phi\nabla^\mu\phi/2\Big] \ , \nonumber 
\end{align}
where $M$ is a mass scale. The first galileon term, $\mathcal{L}_1$, can be removed by a field redefinition and hence ignored; the second galileon term is merely the usual kinetic term for a scalar field. More interestingly, the cubic galileon corresponds to using $\mathcal{L}_2$ and $\mathcal{L}_3$ ($c_4=0$), whilst the quartic employs all contributions ($c_4\neq0$). In fact only ratios of the $c_i$ appear in the galileon field equations, allowing us to fix one $c_i$ parameter arbitrarily. We will use this freedom to fix $c_2=-1$ for both models, so that the field is canonically normalized. 

The choice of the cutoff scale $M$ is an important factor in determining the scales on which the new physics manifests itself. The other important factor in this is the coupling of the galileon field to matter. The interplay between these two factors is nontrivial and highly nonlinear, so that new physics can appear in interesting and unexpected places. For this reason, although the choice $M^3= M_{\rm Pl}H_0^2$ allows the galileon field to have non-trivial dynamics on cosmological scales, and therefore to be a candidate to explain cosmic acceleration, there also exist other astrophysical scales at which additional interesting new phenomena can appear. 

Hereafter we will redefine the scalar field by replacing $\phi/M_{\rm{Pl}} \rightarrow \phi$, rendering it dimensionless. We will also split the scalar field into a homogeneous `background' piece and a perturbation via $\phi=\bar\phi+\varphi$. Note that $\varphi$ is \textit{not} a linear perturbation; it is the deviation of the scalar field from its homogeneous component, i.e. a full nonlinear perturbation.

Before proceeding to our calculation, let us comment briefly on our choice of gravity models here. As mentioned in the introduction, the recent observations of GW170817 and its electromagnetic counterparts \cite{LIGO2017}  constrain the speed of gravitational waves to differ from $c$ by less than one part in  $10^{15}$, which strongly disfavours the quartic (and quintic) galileon models  \cite{Baker2017,Sakstein2017,Zuma2017,Creminelli2017}. Meanwhile, the cubic galileon, when required to drive cosmic acceleration, predicts a negative sign for the Integrated Sachs Wolfe effect (note this has a positive sign in $\Lambda$CDM); this feature is disfavoured by cross-correlation analyses of galaxy surveys and CMB observations \cite{Renk2017}. Hence galileons are no longer a leading candidate for a viable extension of GR for explaining cosmic acceleration.

Despite this, galileons remain a useful toy model for a universe that expands almost identically to $\Lambda$CDM, but has interesting fifth-force phenomenology at the nonlinear level. Their basic features -- a single new scalar degree of freedom, second-order equations of motion, derivative couplings and a handful of free parameters -- are shared by most of the theories of interest on astrophysical scales. Furthermore, terms similar to the cubic and quartic galileons used here appear in the surviving specialised Horndeski, Beyond Horndeski and DHOST theories -- see point 3 of our list in the introduction. As such, we will exploit galileons here to lay out our void lensing methodology, which can be applied to other gravity models (or indeed as a `litmus test' for GR) in future.

\subsection{Galileon Background Dynamics}
\label{subsec:background}
In what follows, we will impose the usual requirement that the galileon field drives cosmic acceleration by setting $M^3 = M_{\rm Pl}H_0^2$. Doing this will allow us to determine the void lensing signal of the most popular models in the recent literature. As explained above, future studies will likely focus on models with heavier mass scales. The Friedmann equation of galileon gravity is (considering the late-time universe, where only pressureless matter and the galileon field are relevant):
\begin{align}\label{eq:friedmann}
3H^2  &= 8\pi G\bar{\rho}_m - \frac{1}{2}{\dot{\bar\phi}}^2 + 6\frac{c_3}{H_0^2}H{\dot{\bar\phi}}^3  + \frac{45}{2}\frac{c_4}{H_0^4}H^2{\dot{\bar\phi}}^4 \end{align}
and the equation of motion for the homogeneous component of the scalar field is:
\begin{align}\label{eq:eom}
-&(\ddot{{\bar{\phi}}}+3H\dot{{\bar{\phi}}}) + \frac{c_3}{H_0^2}\left(12H\dot{{\bar{\phi}}}\ddot{{\bar{\phi}}}+6\dot{H}\dot{{\bar{\phi}}}^2+18H^2\dot{{\bar{\phi}}}^2\right)\nonumber \\
+& \frac{c_4}{H_0^4}\left(54H^2\dot{{\bar{\phi}}}^2\ddot{{\bar{\phi}}}+36\dot{H}H\dot{{\bar{\phi}}}^3+54H^3\dot{{\bar{\phi}}}^3\right) =0 \ ,
\end{align}
where overdots denote physical time derivatives.

\subsection{The Tracker Ansatz}
\label{subsec:tracker}
\noindent The simultaneous solution of eqs.(\ref{eq:friedmann}) and (\ref{eq:eom}) is greatly simplified by the use of the \textbf{tracker ansatz}. This is an approximation for the evolution of the homogeneous component of the galileon field, and has been shown to hold extremely well for galileon models \cite{deFelice2010,Barreira2013a,Barreira2013b}. The tracker ansatz is defined as follows:
\begin{align}\label{eq:tracker}
H\dot{\bar\phi} = \xi H_0^2 \ ,
\end{align}
where $\xi$ is a dimensionless constant. Indeed, it was shown in \cite{BarreiraPlanck} that any galileon model whose solution for $\dot{\bar\phi}$ does not follow the tracker trajectory by $z \simeq 1$ is inconsistent with \textit{Planck} measurements of the CMB temperature power spectrum. Furthermore, the behaviour of the galileon field at $z>1$ has negligible impact on the CMB, so it is safe to adopt the tracker ansatz for all redshifts.

A further implication of the tracker ansatz is a reduction of the free parameters needed to characterize a galileon model. To see this, we substitute
eq.(\ref{eq:tracker}) into eq.(\ref{eq:friedmann}). This converts the Friedmann equation to a fourth-order polynomial in $H$:
\begin{align}\label{friedmann_tracker}
E^4 &= E^2 \Omega_{M0}a^{-3} - \frac{1}{6}\xi^2 + 2c_3\xi^3 + \frac{15}{2}c_4\xi^4  \ ,
\end{align}
where $E= H/H_0$. Evaluating the above expression at $z=0$ gives:
\begin{align}\label{friedman_tracker_z0}
-\frac{1}{6}\xi^2 + 2c_3\xi^3 + \frac{15}{2}c_4\xi^4 &= 1-\Omega_{M0}  \ .
\end{align}
Similarly, using the tracker ansatz in eq.(\ref{eq:eom}) and evaluating at $z=0$ yields:
\begin{align}\label{eom_tracker}
\left[3-\frac{\dot H}{H^2}\right]\left[-\frac{1}{3}+2c_3\xi+6c_4\xi^2\right]&=0 \ . 
\end{align}
The second square bracket here must vanish, since the first cannot do so for a general background expansion rate. Together, eqs.(\ref{friedman_tracker_z0}) and (\ref{eom_tracker}) give two algebraic constraints relating the parameters $\xi$, $c_3$ and $c_4$. After solving these constraints, the cubic galileon ($c_4=0$) has no free parameters remaining (other than the standard cosmological parameters), whilst the quartic model has one free parameter. These constraints are therefore (our expressions are equivalent to those of \cite{Renk2017}):\vspace{2mm}

\noindent \textbf{Cubic:}
All parameters are fixed by the value of $\Omega_{M0}$:
\begin{align}
\xi&=\sqrt{6(1-\Omega_{M0})}\simeq 2.05 \label{cubic_constraints1}
\\
\quad c_3 &=\frac{1}{6\xi}\simeq 0.08 \ .\label{cubic_constraints2}
\end{align}
\textbf{Quartic:}
Here we take the one free parameter to be $\xi$, the tracker constant of proportionality. Then we have:
\begin{align}
c_3 &= -\frac{2}{\xi^3}\left(1-\Omega_{M0}\right)+\frac{1}{2\xi} \label{quartic_constraints1}
\\
c_4 &= \frac{2}{3\xi^4}\left(1-\Omega_{M0}\right)-\frac{1}{9\xi^2} \ .\label{quartic_constraints2}
\end{align}

We will use these constraints to fix some of our galileon model parameters. For a given value of $\xi$, eq.(\ref{friedmann_tracker}) becomes a quadratic equation for $E^2$. Choosing the solution that gives real $H$, the Hubble rate is then given by:
\begin{align}\label{eq:tracker_H}
E(a)^2 &= \frac{1}{2}\left[\Omega_{M0}a^{-3} \right.   \left. + \sqrt{\Omega_{M0}^2a^{-6} + 4(1-\Omega_{M0} )}\right] \ .
\end{align} 
Compared to eqs.(\ref{eq:friedmann}) and (\ref{eq:eom}), this represents a remarkable simplification of the background evolution for our models. 

\subsection{Quasistaticity \& Pathologies}
\label{subsec:quasistatic}

Whilst the tracker ansatz is strongly supported by measurements of the cosmological background expansion rate, another commonly-used assumption is surrounded by a higher degree of uncertainty. The quasistatic (QS) approximation is a statement about the relative timescales that characterise cosmological structure formation. It states that cosmological structures evolve on approximately Hubble timescales, and hence that the time derivatives of linear cosmological perturbations are expected to be suppressed compared to their spatial derivatives (which vary on significantly sub-Hubble scales). 

In addition to the metric potentials, the QS approximation is usually assumed to also apply to the perturbations of any new fields present in a modified gravity theory, i.e. $|\partial_i\varphi|\gg|\dot\varphi|$ for the galileon field.  This treatment has been shown to hold on scales larger than the sound horizon of the scalar field \cite{Sawicki2015} in several theories \cite{Noller2014,Bose2015} (though see \cite{Burrage2017} for a study of structure formation in galileon gravity \textit{without} the QS approximation). However, these results do not guarantee that the QS approximation holds for all theories. Indeed, the QS assumption has been called into question by the discovery that galileons display pathologies in certain regimes of cosmological structure formation. Specifically,~\cite{Winther:2015pta,Barreira:2013eea} have found that the solution for the galileon field profile becomes imaginary inside voids of a certain depth and redshift. The region of imaginary solutions does not span the entire void, but occupies a central region whose radius varies with void depth and redshift. We will meet these pathologies in our own results in \S\ref{section:theory results}.

Whether these unphysical solutions signal a real breakdown of galileon theories, or are an artifact of the breakdown of the QS approximation, remains to be seen. The full set of galileon field equations, free of the QS approximation, is a complex set of coupled, nonlinear partial differential equations with derivative interactions, and requires a thorough numerical treatment beyond the scope of this paper. However, in what follows we will at least attempt to explain why these pathologies occur, and pursue a careful delineation of the parameter regime (in void depth vs. radius) in which they arise.

\section{Void Calculation}
\label{section:voidcalc}

\subsection{Density Profiles and Void Sizes}
\label{subsec:density}
In the next subsection we will calculate the gravitational fields associated with a spherical underdensity. Although any given void is likely to be non-spherical, the averaged density profile of all voids in a given sample (of sufficient size) should be spherical to a very good approximation. In this paper we will make use of two different density profiles that have been explored in the literature. The first of these profiles is a simple cubic fit, whilst the second profile has greater flexibility; it has additional parameters controlling a compensation ridge around the void, and hence can be used for a wider range of void sizes (smaller voids of around $10-30$ Mpc/h tend to be compensated by an external ridge, whilst larger ones with $R_V \gtrsim 50$ Mpc/h do not \cite{Hamaus2014}).

One may question whether it is valid to use void profiles originally derived from GR simulations, or fitted to survey data assuming GR, for our work here. Given a measured galaxy profile, obtaining the matter profile responsible for lensing may require some modelling in a GR context. Assuming that galaxies are linearly biased, which is likely to be valid at scales $\sim R_v/2$ \cite{Fanginprep}, the GR matter density profile is simply a rescaled version of the galaxy profile. The scaling factor is set by the galaxy bias, which is known from the galaxy auto-correlation function. On scales closer to the void center, a GR based mock catalog is required to model the relation of the matter profile to galaxies. This is an area of ongoing research \cite{Pollina2018,Fanginprep}, and it remains to be determined fully whether such relations are significantly affected by deviations from GR.  

However, prior work on the cubic galileon \cite{Barreira2015} (as well as on $f(R)$ theories \cite{Cai2014,Cai2015}) indicates that the density profiles of voids are not significantly altered in these theories. These works found that any changes were at the level of a few percent or less (we describe the generic reasons to expect small effects in the density profile in \S IV A). Such effects are small compared to the errorbars of the SDSS void lensing data we use in this paper (Fig.~\ref{fig:SDSS}), but will need to be accounted for with future measurements. For the present, we proceed to use the GR-based density profiles for tests of galileon gravity; as we will see in \S\ref{section:theory results}, the impact of modified gravity on the lensing signal can still be substantial.

The form of the simpler profile is:
\begin{align}
\label{simple_profile}
\delta(R) &=\delta_v\left[1-R^3\right] \quad\quad &0 < R < 1\\
\delta(R) &=0 &R>1 \ ,
\end{align}
where $\delta = \delta\rho/\bar\rho$ is the density fluctuation about the mean matter density $\bar\rho$, and $R=r/R_V$ is the radial coordinate in units of the void radius $R_V$. The single parameter $\delta_v$ then describes the maximum central density contrast of the void. Introduced in \cite{Clampitt2015}, this is a simpler form of the cubic profile originally studied in~\cite{Lavaux:2011yh,Krause:2012aq}.

The flexible profile we use is \cite{Hamaus2014}:
\begin{align}
\label{complex_profile}
\delta(R)&=\delta_v\left[ \frac{1-(R/s_1)^\alpha)}{1+(R/s_2)^\beta}\right] \ ,
\end{align}
where again $\delta_v$ is the maximum density contrast. Variants of this profile have been explored in \cite{Barreira2015, Sanchez2017}, where it was found to provide a good fit to voids in simulations of alternative gravity theories (as well as in $\Lambda$CDM). Clearly the disadvantage of this profile is a larger number of parameters to be fitted. An in-depth study of the effects of all five parameters can be found in \cite{Barreira2015}, and we will keep most of these fixed at their best-fit values. We will, however, study the effects of ridge height on the tangential shear signal at large radii in \S\ref{subsec:profiles}. The fiducial values of the parameters in eq.(\ref{complex_profile}) are:
\begin{align}
\alpha&=3 & s_1&= 0.9 \nonumber\\
\beta&=7 & s_2&=1.1 \ .
\label{complex_params}
\end{align}

Generically, one may expect the void size function and central underdensity distrbution to differ in alternative theories of gravity (although, as we discussed above, the shapes of the profiles seem to be largely unaffected). The additional force component mediated by the scalar field typically acts to evacuate matter from a void more quickly\footnote{In the quartic galileon model, for some values of the tracker parameter $\xi$, the scalar can actually act to suppress gravitational forces.}, enhancing the number of strongly under-dense voids. A full treatment of the issue requires a study of voids in numerical simulations of modified gravity, which is beyond the scope of this paper. In \S\ref{subsec:profiles} we will address this issue in part, by comparing the probability distribution functions of the central underdensity in two models.

\subsection{Semi-linearised Field Equations}
\label{subsec:field_eqs}
We now proceed to calculate the behaviour of the gravitational and galileon fields over voids with the aforementioned density profiles. We keep the galileon field and matter perturbations fully nonlinear, but the perturbations to the gravitational metric remain small. We describe them using the following metric convention (in conformal Newtonian gauge):
\begin{align}
\label{lineel}
ds^2&=-\left(1+2\Psi\right)dt^2+\left(1-2\Phi\right)\left(dr^2+r^2d\Omega^2\right) \ .
\end{align}
In what follows we will rescale the radial coordinate as $\chi = a H_0 r$. The full field equations are lengthy and given in \cite{Barreira2013b}; we will not reproduce them here. After making the quasistatic approximation and integrating once over the radial coordinate, the equations simplify to two equations for the metric perturbations:
\begin{align}
\label{eq:poisson}\frac{\Phi,_{\chi}}{\chi} &=\frac{1}{A_4} \left[\Omega_{m0}a^{-3}\frac{\delta M}{r^3}  + A_1\left(\frac{\varphi,_{\chi}}{\chi}\right) + A_2\left(\frac{\varphi_{\chi}}{\chi}\right)^2 \right] \ , \\
\label{eq:slip}\frac{\Psi,_{\chi}}{\chi} &=\frac{1}{B_3} \left[B_0\left(\frac{\Phi,_{\chi}}{\chi}\right) + B_1\left(\frac{\varphi,_{\chi}}{\chi}\right) + B_2\left(\frac{\varphi,_{\chi}}{\chi}\right)^2\right] \ ,
\end{align}
where subscript commas indicate derivatives. These are supplemented by the equation of motion for the scalar field:
\begin{align}
\label{eq:eom_sph} 
0&= C_1\frac{\varphi,_{\chi}}{\chi} + C_2\left(\frac{\varphi,_{\chi}}{\chi}\right)^2 + C_3\left(\frac{\varphi,_{\chi}}{\chi}\right)^3  + C_4\frac{\Phi,_{\chi}}{\chi}\nonumber\\
& + C_5\frac{\Psi,_{\chi}}{\chi}  + C_6\frac{\varphi,_{\chi}}{\chi}\frac{\Phi,_{\chi}}{\chi} + C_7\frac{\varphi,_{\chi}}{\chi}\frac{\Psi,_{\chi}}{\chi}  \ .
\end{align}
Here, the quantities $A_i$, $B_i$ and $C_i$, defined in the appendix, are functions of the background cosmology, and hence depend only on time.

Using eqs.(\ref{eq:poisson}) and (\ref{eq:slip}) to eliminate derivatives of $\Phi$ and $\Psi$, eq.(\ref{eq:eom_sph}) simplifies to a third-order algebraic polynomial for $\varphi,_{\chi}/\chi$:
\begin{align}
\label{eom_phi}
0 &=  \eta_{01}\frac{\delta M}{r^3} + \left(\eta_{11}\delta M/r^3 + \eta_{10}\right)\left[\frac{\varphi,_{\chi}}{\chi}\right]  + \eta_{20}\left[\frac{\varphi,_{\chi}}{\chi}\right]^2 \nonumber\\
&+ \eta_{30}\left[\frac{\varphi,_{\chi}}{\chi}\right]^3 \ .
\end{align}
The functions $\eta_i$, defined in the appendix, are linear combinations of the $A_i$, $B_i$ and $C_i$. The true polynomial order of eq.(\ref{eom_phi}) depends on which of the galileon terms are present. In the cubic galileon one has $\eta_{30} = \eta_{11}=0$, and eq.(\ref{eom_phi}) is then a quadratic equation in $\varphi_\chi/\chi$. For the quartic galileon, the equation is cubic in $\varphi_\chi/\chi$. Hence, in both cubic and quartic models, there are multiple branches of solutions.

This raises the question of which branch of solutions is the physically realised one? The standard protocol here is to select the branch for which $\varphi_{,\chi}/\chi\rightarrow 0$ as $\delta M/r^3 \rightarrow 0$, on the grounds that there cannot be a self-sustained field configuration for $\varphi$ in the absence of any mass fluctuation. We will adopt this convention in the absence of any pathologies, which are signaled by the onset of complex solutions for $\varphi_{,\chi}/\chi$. If the dynamics drive the system into a pathological region, we will then instead choose the remaining real solution, if one exists (we will see in \S\ref{subsec:path} that there is no ambiguity in this choice). If all solution branches becomes imaginary in a particular region of void parameter space and radius, then we deem the theory and/or approximations used to be pathological (and therefore meaningless) there. We will set such regions to zero values in our plots.

Having selected the appropriate solution branch for $\varphi_{,\chi}/\chi$, we use this in eqs.(\ref{eq:poisson}) and (\ref{eq:slip}) to calculate derivatives of metric potentials. We can also straightforwardly take a second radial derivative of (\ref{eq:poisson}) and (\ref{eq:slip}), although we will not show the resulting lengthy expressions here. We then have in hand all the quantities necessary to calculate a void lensing signal.

\subsection{Lensing Integrals}
\label{subsec:lensing}
The lensing convergence of a void, $\kappa$, is obtained by integrating the sum of the second derivatives of the metric potentials along the line of sight to the void:
\begin{align}\label{eq:k-def}
\kappa &= \frac{1}{4\pi G \Sigma_{\rm{crit}}}\,\frac{1}{2} \int \nabla^2\left(\Phi+\Psi\right) {\rm d}l \ ,\\
&= \frac{1}{4\pi G \Sigma_{\rm{crit}}}\,\frac{a^2H_0^2}{2} \int\left[ \Phi_{,\chi\chi}+\Psi_{,\chi\chi}+\frac{2}{\chi}\left(\Phi_{,\chi}+\Psi_{,\chi}\right) \right]{\rm d}l  \label{kdef2}
\end{align}
where, in the second line, we have written the derivatives in our scaled radial coordinate, $\chi = a H_0 r$. The critical density of a lensing system, $\Sigma_c$, is defined in terms of the angular diameter distance $D_A(z)$ as:
\be
\label{uio}
\Sigma_{\rm crit} (\zl, \zs) = \frac{c^2}{4\pi G} \frac{D_A(\zs) (1+\zl)^{-2}}{D_A(\zl) D_A(\zl,\zs)} \ ,
\ee
where $z_s$ and $z_L$ are the redshifts of the source and lens (in this case, the void) respectively, and the $(1+\zl)^{-2}$ factor is due to our use of comoving coordinates. We set $\Sigma_{\rm crit}^{-1}(\zl,\zs)=0$ for $\zs<\zl$ in our computations. 

The observable quantity of interest here is the differential surface mass density,
\begin{align}
\Delta\Sigma(r)&=\Sigma_c\,(\bar{\kappa}-\kappa) \ ,
\label{kat}
\end{align}
where $\bar{\kappa}$ is the mean convergence as a function of radius from the centre of the void, given by:
\begin{align}
\label{kappabar}
\bar{\kappa}&=\frac{2R_V}{R^2}\int_0^R\,y\,\kappa(y)\,dy \ .
\end{align}
The distances to the source galaxies and void are not needed for the calculation of an individual void lensing profile; that said, the source and void redshift distributions \textit{are} needed when calculating the stacked void lensing profile for an entire survey. Note that $\Delta \Sigma$ is in fact the projected effective density profile of the lensing void\footnote{To see this, note that the integrand of eq.(\ref{eq:k-def}) can be replaced using eq.(\ref{eq:poisson}).}. However, eq.(\ref{kat}) can equivalently be written as $\Delta\Sigma = \Sigma_{\rm crit}\times \gamma_t $, where $\gamma_t =\bar{\kappa}-\kappa$ is the lensing tangential shear, so $\Delta\Sigma$ also carries the modifications to the lensing signal. We will express most of our results in terms of $\Delta \Sigma$, since it is a convenient variable for comparing theory and measurements.

\section{Results}
\label{section:theory results}
In this section we present the results of our calculation for the void shear profiles, showing the impacts both of deviations from GR and of variations of the void density profile. We also examine the reported onset of pathological behaviour for some parts of the void parameter space. All calculations in this section use best-fit values for the standard cosmological parameters from the \textit{Planck} results \cite{Planck2016}.

\subsection{Gravity Models}
\label{subsec:models}
Fig.~\ref{fig:models} shows the expected tangential shear profiles for voids governed by General Relativity and by the cubic and quartic galileon models, for the void density profile of eq.(\ref{complex_profile}). Based on the discussion of \S\ref{subsec:density}, we use the same input void density profile for all three gravity models (upper panel), and study their differing lensing predictions (lower panel).

\begin{figure}
\centering
\hspace{-6mm}
\subfigure{\label{fig:models}\includegraphics[scale=0.47]{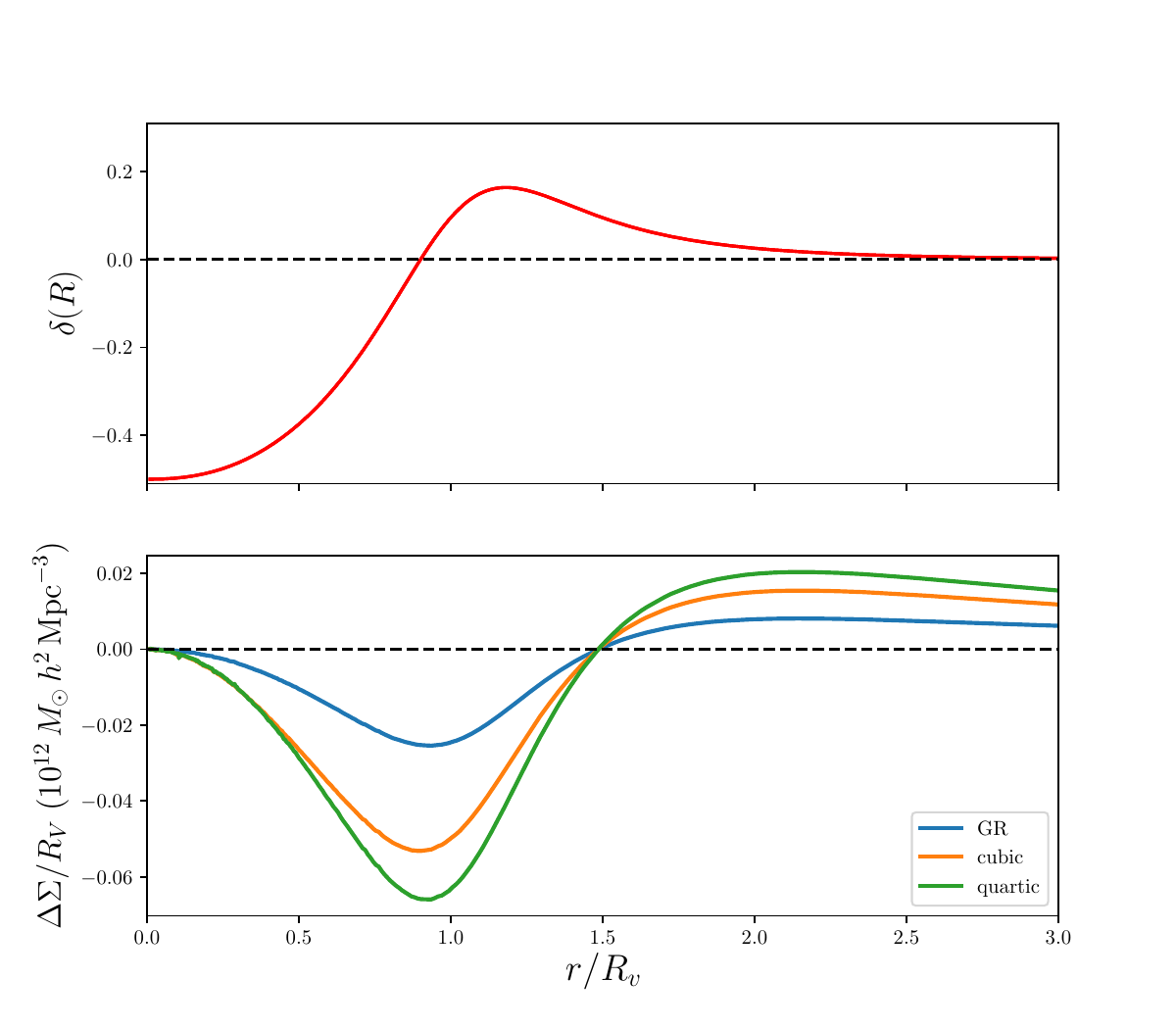}}%
\caption{\textit{Upper panel:} the void density profile of eq.(\ref{complex_profile}), shown here with a central depth $\delta_v = -0.5$ and the fiducial parameters of eq.(\ref{complex_params}). \textit{Lower panel:} corresponding tangential shear profiles in GR, the cubic galileon and the quartic galileon gravity theories. Recall (from \S\ref{subsec:tracker}) that after applying the tracker ansatz the cubic galileon has no free parameters, whilst the quartic galileon has one; we take this here to be $\xi=1.9$. This figure is shown at $z=0$.}
\label{fig:models}
\end{figure}

We plot the quantity $\Delta\Sigma/R_V$, which -- for a fixed void profile at least -- is independent of void size, since $\Delta\Sigma$ scales linearly with void radius (see eq.\ref{kappabar})\footnote{The subtlety here being that, in reality, small and large voids tend to have slightly different profiles, see \S\ref{subsec:profiles}.}. 

It is clear from Fig.~\ref{fig:models} that the cubic galileon model, having no free parameters after imposition of the tracker ansatz, gives rise to significant deviations from the GR tangential shear signal. There is a factor of $\sim 2$ boost in the lensing signal compared to the GR predictions, at all radii. 

As described in the introduction, modified gravity theories can alter the lensing signal via two channels: i) by contributing to the the effective energy density on the RHS of the Poisson equation, and ii) by creating effective anisotropic stress such that $\Phi\neq\Psi$. Using eq.(\ref{eq:slip}) and the expressions in the appendix, one can see that the cubic galileon does not generate any effective anisotropic stress. Consequently, all the deviations between the GR and cubic galileon curve in Fig.~\ref{fig:models} must arise from the effective energy density of the galileon field.  

Initially it may seem surprising that the galileon field can have such a substantial effect on the lensing profile, whilst its effects on the matter distribution within the void are much smaller (see \S\ref{subsec:density}). The reason underlying this is the relative evolutionary timescales of the matter distribution and the scalar field profile. Since the galileon field is designed to drive cosmic acceleration (at least in the model considered here), it only becomes a significant fraction of the energy budget of the universe at late times, $z<1$ typicall (see \cite{Lue2004} for the evolution of spherical perturbations in a comparable gravity model). The void density profile has largely been determined before these redshifts are reached. A related point concerns the galaxy profile. Since galaxies form in the high density environments of halos, the relation of galaxies to matter is likely to remain close to that in GR, except possibly for lower mass halos. We do not consider the details of the galaxy-matter density fields any further. 

\begin{figure}
\centering
\hspace{-5mm}
\subfigure{\includegraphics[scale=0.47]{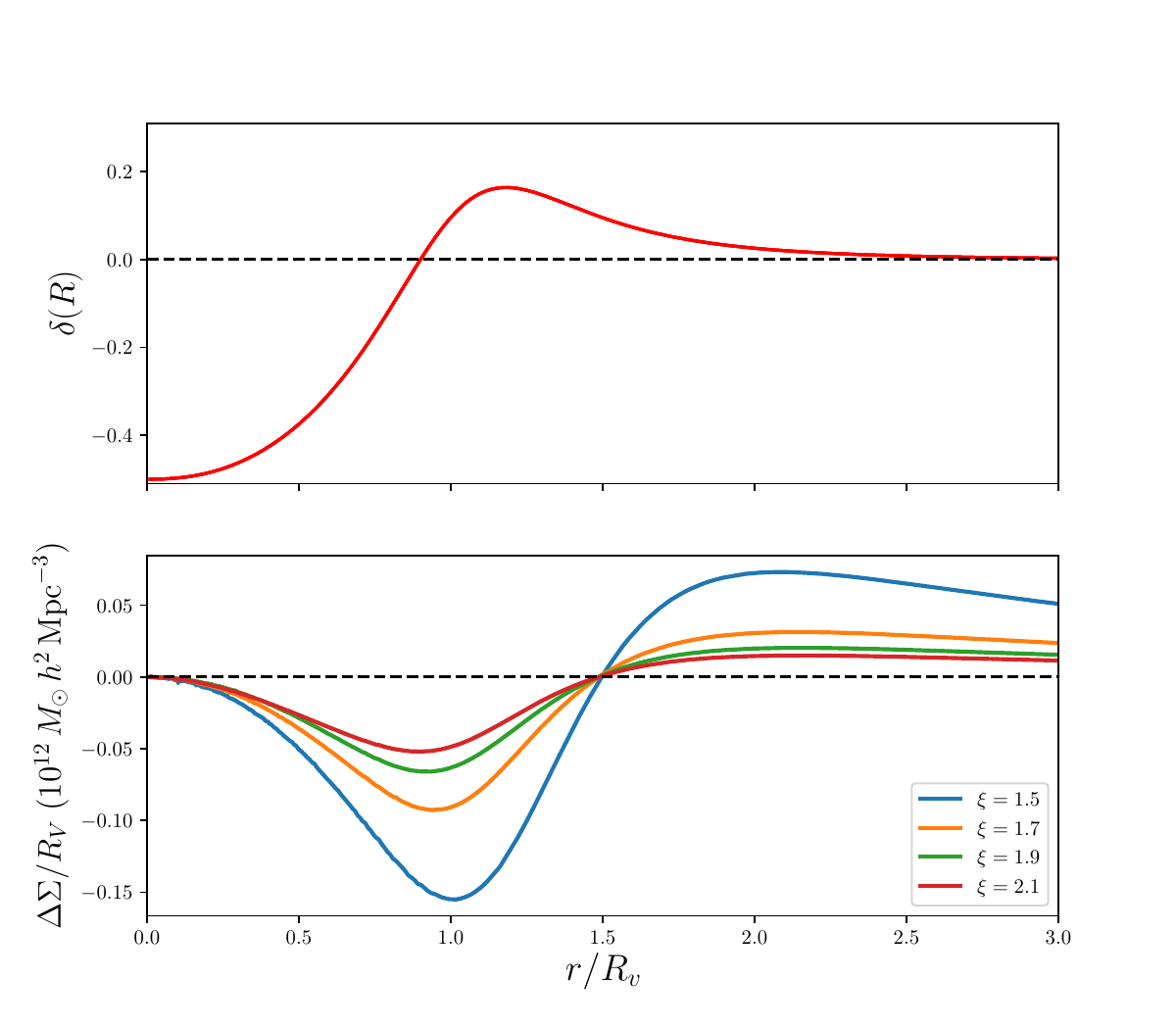}}%
\caption{The effect of varying $\xi$ in the quartic galileon model; in the cubic galileon model this parameter is fixed to $\xi\sim 2.1$. The void density profile and parameters are the same as in Fig.~\ref{fig:models}.}
\label{fig:xi}
\end{figure}

The quartic galileon is an example of a theory which can modify lensing via both of the channels above. Despite this, the quartic curve in Fig.~\ref{fig:models} shows only a moderate further enhancement relative to the cubic galileon prediction. This is due to the effect of the constraints in eqs.(\ref{cubic_constraints1}-\ref{quartic_constraints2}), which fix the value of $c_4$ used to be small relative to $c_3$. Quantitatively, using $\xi=2.05$ in the quartic galileon gives $c_3\simeq0.08$, similar to the cubic value, but $c_4\simeq -1.5\times 10^{-5}$ (and for $\xi=1.9$ as shown in Fig.~\ref{fig:models}, $c_3\simeq 0.06$ and $c_4\simeq 5\times10^{-3}$ ). This explains why the variation of the lensing amplitude between the cubic and quartic models is less significant that their shared substantial difference from GR. However, Fig.~\ref{fig:xi} shows that the tangential shear profile of the quartic galileon is quite sensitive to small variations away from $\xi=1.9$, particularly for $\xi<1.9$.

Note from Figs.~\ref{fig:models} and \ref{fig:xi} that, despite significant variation around the void radius and at a few radii out ($r/R_V\sim2-3$), the null of the tangential shear remains fixed at $r\sim 1.5\, R_v$ in all cases. The reason for this is as follows: the void density profile determines the radius at which the void is exactly compensated, i.e. $\delta M(<r)\rightarrow 0$. In \S\ref{subsec:field_eqs} we selected the physical branch of solutions such that $\phi_{,\chi}/\chi\rightarrow 0$ at the same radius. Since the void density profile used is the same for all gravity models (see discussion in \S\ref{subsec:density}), the potential derivatives (eqs.~\ref{eq:poisson} and \ref{eq:slip}) vanish at the same radius for all gravity models. Via eq.(\ref{kdef2}), this then ensures that the null of $\Delta\Sigma$ is unchanged by variations of the gravity model. This property should hold  true for any model of gravity that does not appreciably impact the void density profile.

\subsection{Void Profiles}
\label{subsec:profiles}
\begin{figure}
\centering
\resizebox{95mm}{!}{\includegraphics{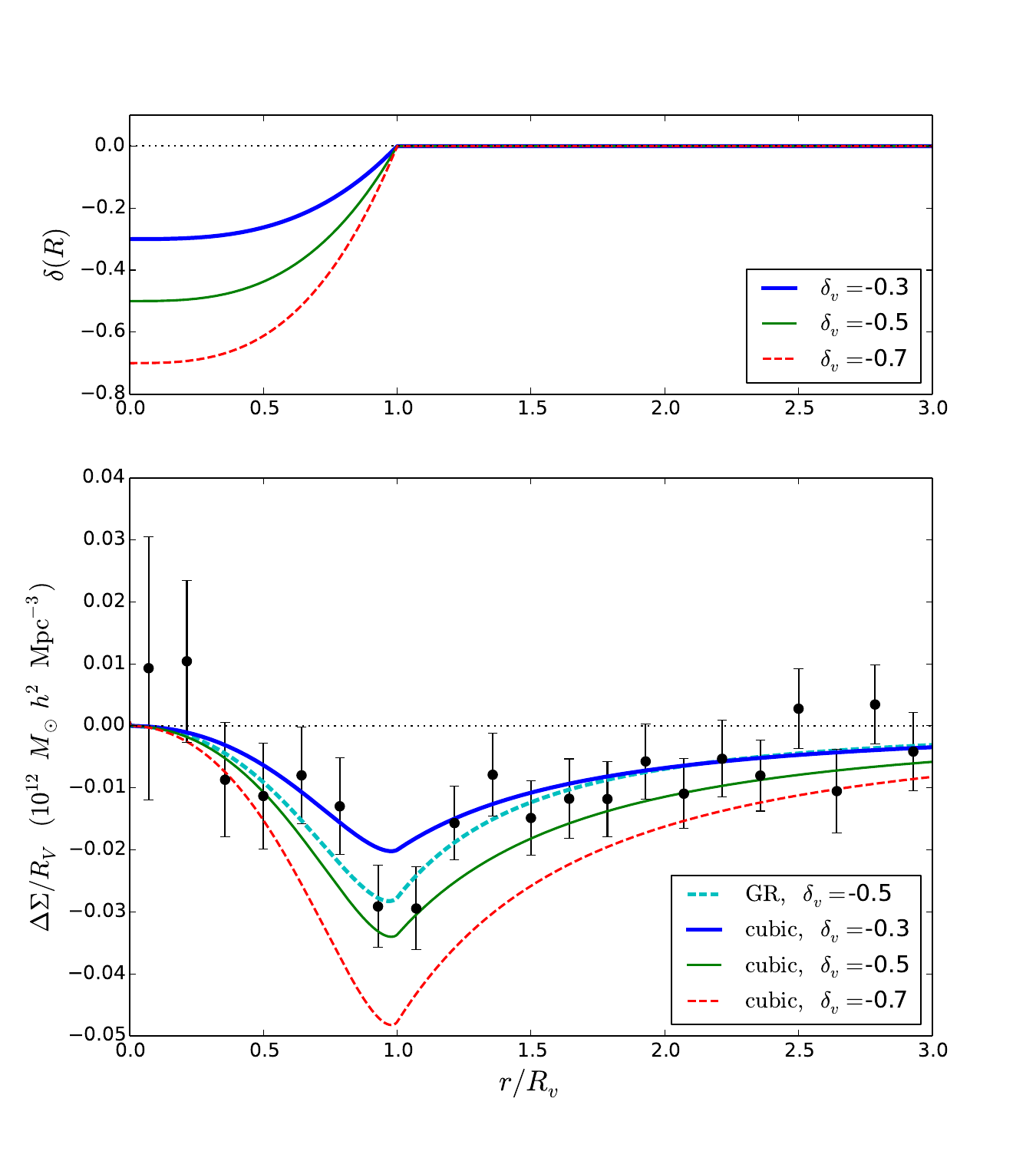}}
\caption{\textit{Upper panel:} the cubic density profile of eq.(\ref{simple_profile}). \textit{Lower panel:} the corresponding shear profiles in GR and the cubic galileon model. Overplotted are SDSS void lensing measurements.}
\label{fig:SDSS}
\end{figure}

In \S\ref{subsec:density} we introduced two void density profiles: a simple cubic fit, and a compensated ridge profile. For ease of comparison, most of the figures in this paper employ the latter profile. In Fig.~\ref{fig:SDSS} we show the corresponding density and tangential shear profile for the cubic fit, together with data obtained from voids identified by \cite{Clampitt2015} in the SDSS DR7-Full LRG catalog of \cite{Kazin2010}. As reported in \cite{Clampitt2015}, a void of central depth $\delta_v\simeq -0.5$ provides a good fit to the data in GR. 

In the lower panel we further show a corresponding set of tangential shear profiles in the cubic galileon model. It is clear that the cubic galileon produces a higher amplitude lensing signal than in GR (this can be seen, for example, by comparing the two curves with $\delta_v=-0.5$), and that this enhancement persists out to distances well beyond the void radius. 

We will use the void lensing data and covariance (as per the methods of \cite{Clampitt2015}) to obtain the posterior probability of $\delta_v$ for the cubic galileon. It is easy to see by eye from Fig. 3 that a cubic galileon model with $\delta_v \sim -0.4$ provides a good fit to the SDSS data points (compared to  $\delta_v \sim -0.5$ in GR). Hence we expect there to be some shift in the distribution of $\delta_v$ for different gravity models.

\begin{center}
\begin{figure}
\hspace{-7mm}
\resizebox{95mm}{!}{\includegraphics{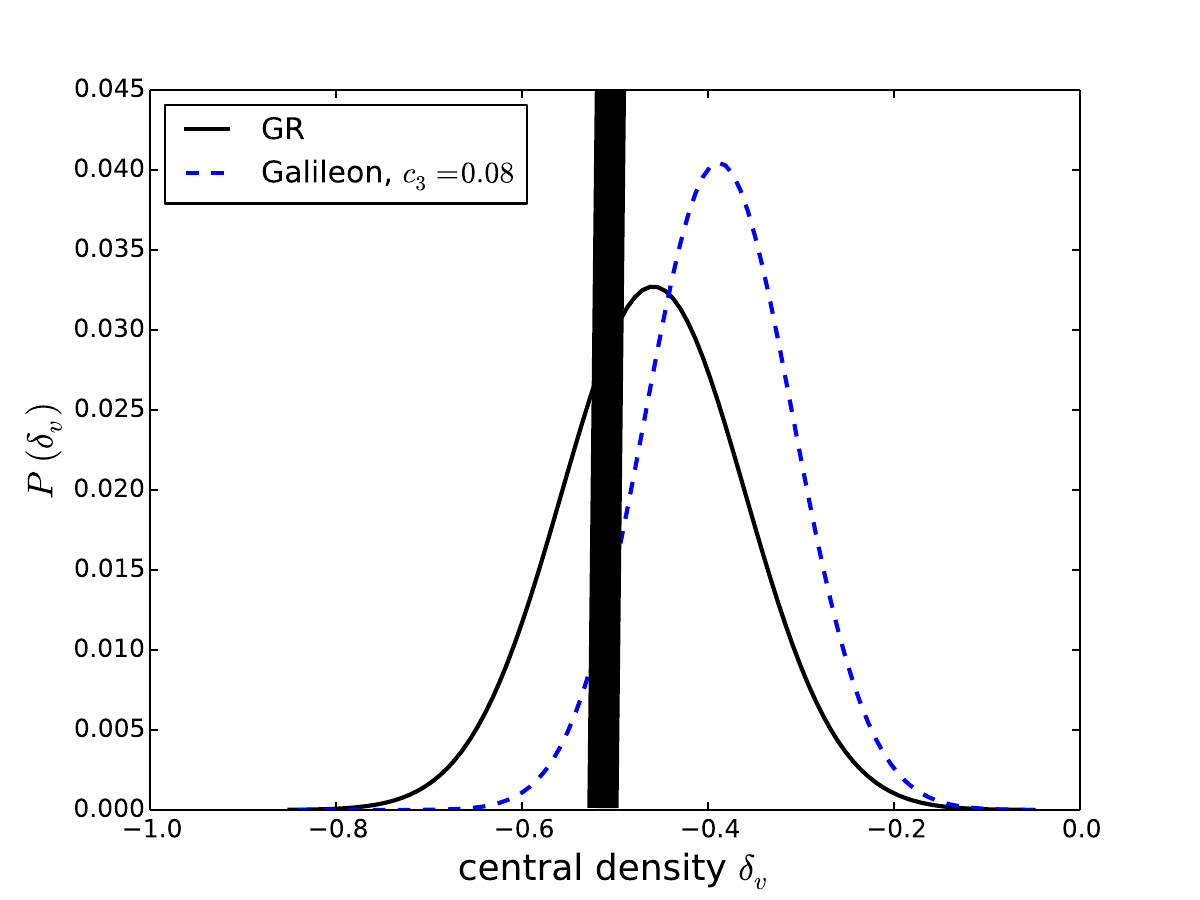}}
\caption{The best-fit values of the void central density parameter $\delta_v$, as inferred from the lensing profiles of the SDSS DR7-Full LRG catalog. The black curve is for GR, while the dashed, blue curve assumes the cubic galileon model. We assume that the relation of galaxies to mass is not altered in modified gravity, based on simulation studies (Barreira et al 2015).}
\label{fig:density}
\end{figure}
\end{center}

Fig.~\ref{fig:density} shows our results. These confirm that in the cubic galileon, the best-fit $\delta_v$ shifts towards more shallow voids than in GR. This makes sense, since a given value of $\delta_v$ produces a larger lensing signal in the cubic galileon than it does in GR -- see Fig.~\ref{fig:models}. Reversing the argument, a given lensing signal maps back onto a shallower void in the cubic galileon than it does in GR. 

We will not pursue here a more detailed comparison of our theoretical predictions to current observations. This is due to significant uncertainties in obtaining the mass profile of voids from their galaxy tracer profiles, in gravity models outside of GR. A careful analysis using mock catalogs would be required to obtain these constraints; we leave this to a future work. 

\begin{center}
\begin{figure}
\hspace{-5mm}
\resizebox{95mm}{!}{\includegraphics{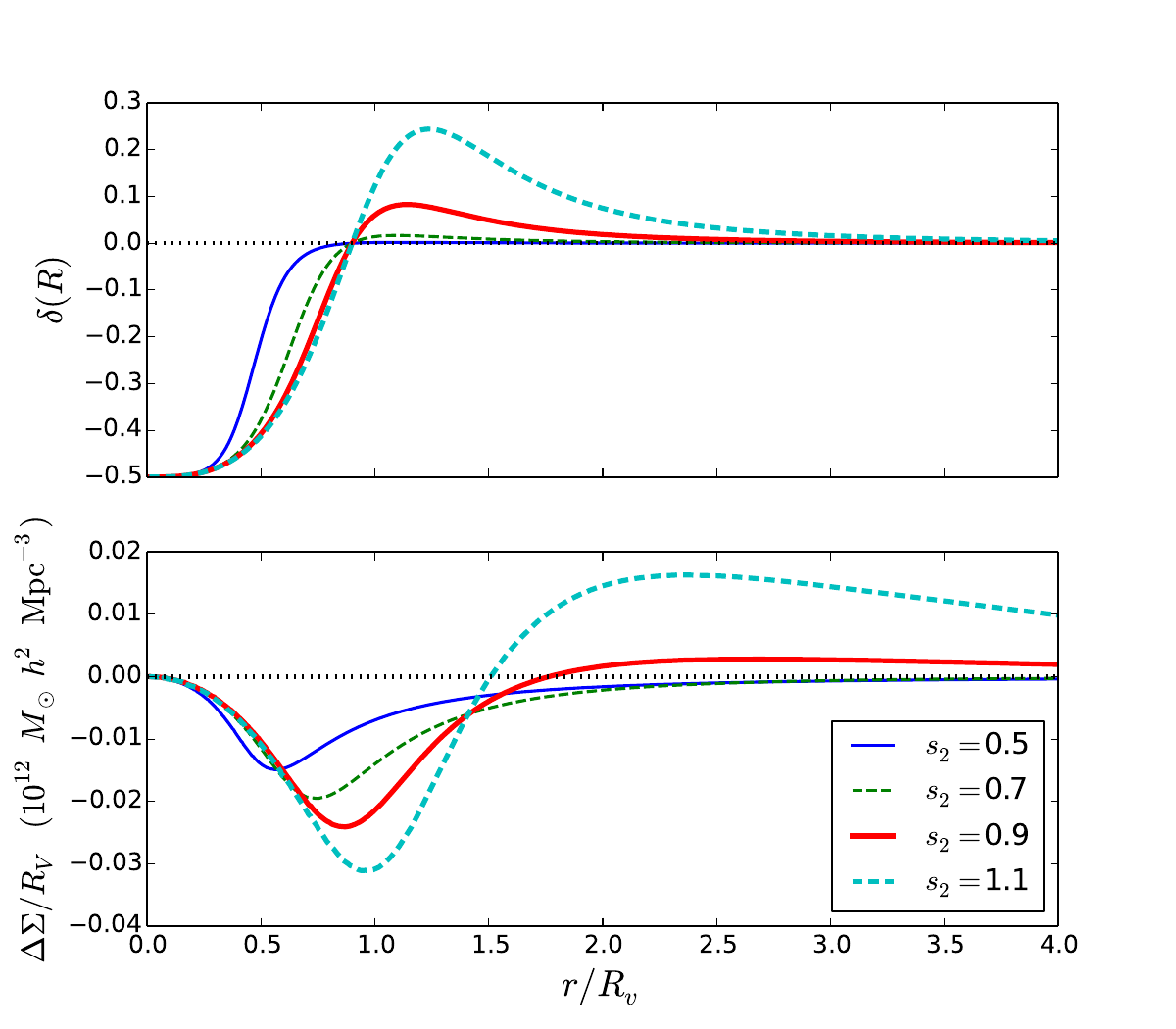}}
\caption{\textit{Upper panel:} realisations of the density profile in eq.(\ref{complex_profile}), with varying values of $s_2$. This parameter mainly controls the height of the compensating ridge. \textit{Lower panel:} corresponding shear profiles in GR. The lensing amplitude and the location of the minimum of the signal are affected substantially by the ridge height.}
\label{fig:s2}
\end{figure}
\end{center}

In Fig.~\ref{fig:s2} we return to the flexible density profile and regular GR, and demonstrate the variation in tangential shear induced by the presence of the compensation ridge. In reality, ridge height is inversely correlated with void radius, and hence correlated with void depth (i.e. small, deep voids have pronounced ridges, see figure 1 of \cite{Hamaus2014}). Here we wish to disentangle the effects of void depth and ridge height, which we can do by varying the parameter $s_2$ of eq.(\ref{complex_profile}) but keeping $\delta_v$ fixed. Although the other parameters in eq.(\ref{complex_params}) can also affect the height of this ridge, $s_2$ has the most significant effect (see Fig. 8 of \cite{Barreira2015}). We see from Fig.~\ref{fig:s2} that the the presence of an exterior ridge significantly boosts the shear profile at large radii, and shifts the minimum of the shear profile to larger radii. 

Whilst we have not done a detailed analysis here, it is evident (even by eye) that the SDSS measurements in Fig.~\ref{fig:SDSS} do not support the existence of a compensated ridge, and would rule out large values of $s_2$. With future higher signal-to-noise measurements, one can bin voids by size, since smaller voids have more prominent ridges. Given that deviations from GR tend to boost lensing at large radii, much like a ridge does, ridge features need to be studied in cosmological simulations and carefully accounted for when using lensing measurements to test gravity. A subtlety here is that the choice of void tracer and void finder can impact the analysis \cite{Colberg2008, Nadathur2015c, Sanchez2017,Cautun2018}.

\subsection{The Pathological Regime}
\label{subsec:path}
As alluded to in \S\ref{subsec:quasistatic}, it is known that galileon gravity suffers from pathological solutions under certain conditions. By `pathological', we mean that the solutions for the scalar field radial derivative $\phi_{,\chi}/\chi$, become imaginary, and hence the lensing calculation laid out here breaks down. 

As we have noted, eq.(\ref{eom_phi}) is a polynomial in $\phi_{,\chi}/\chi$ -- the order of which depends on the model --  and hence will generally possess multiple solutions. For the cubic galileon, eq.(\ref{eom_phi}) is a quadratic polynomial in $\phi_{,\chi}/\chi$ with entirely real coefficients, so any complex solutions must form a conjugate pair. Hence both solutions become pathological simultaneously. This is in contrast to the quartic galileon, for which eq.(\ref{eom_phi}) is a cubic polynomial in $\phi_{,\chi}/\chi$. Though the quartic galileon also forms pairs of complex solutions in certain void regimes, we have verified that the physical value of $\phi_{,\chi}/\chi$ (the one which tends to zero as $\delta M/r^3\to 0$) corresponds to the one remaining real solution of the cubic polynomial. Hence, the quartic galileon remains pathology-free inside voids\footnote{We note that the quintic galileon -- which we do not discuss in this paper -- has the unusual feature of displaying pathologies inside \textit{overdense} regions \cite{Barreira2013b}.}.

\begin{figure*}
\begin{center}
\subfigure{\label{fig:s2var}\includegraphics[scale=0.7]{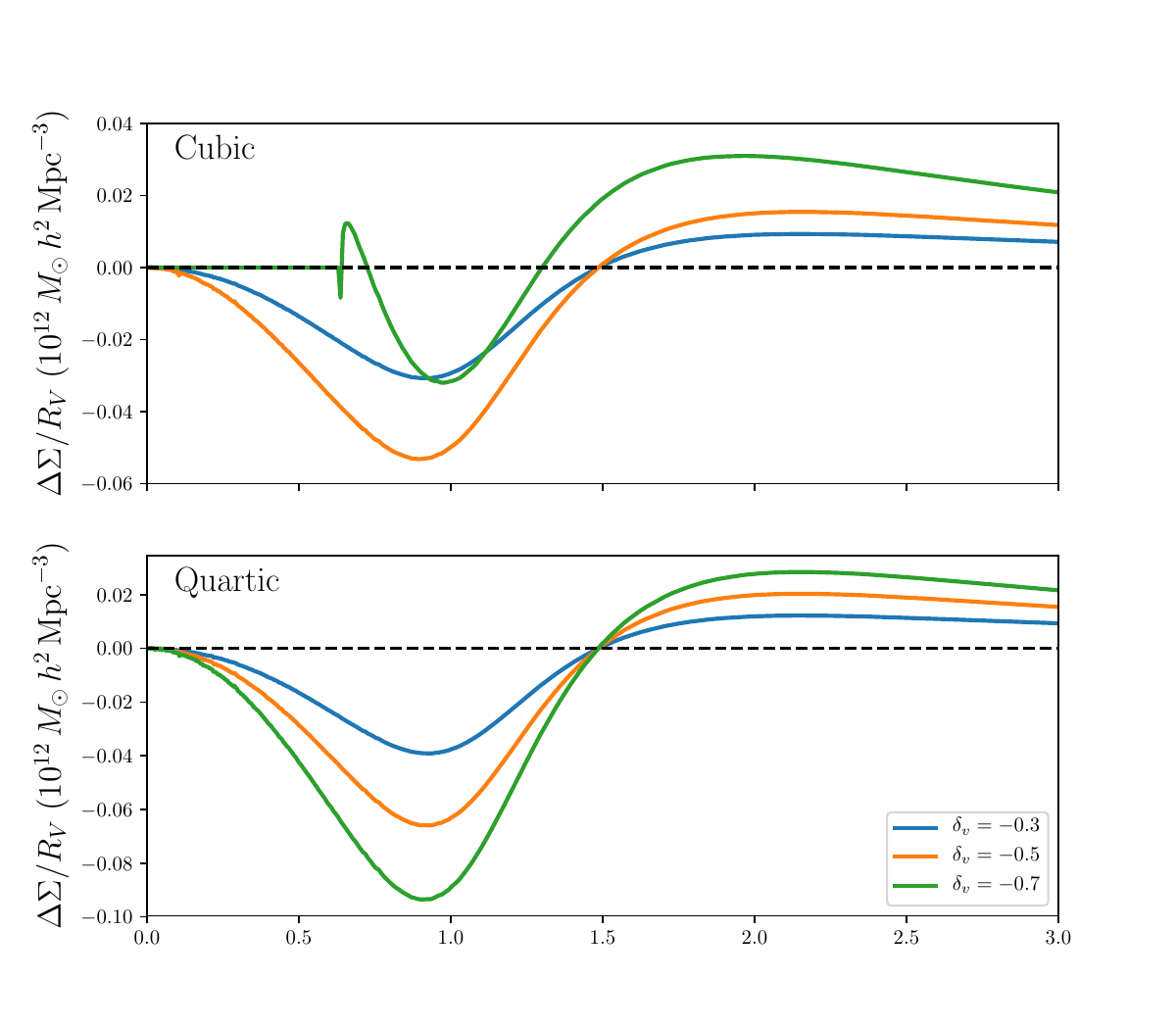}}%
\caption{Variation of lensing tangential shear with void depth and gravity model, for the density profile of eq.(\ref{complex_profile}), shown for $z=0$, $R_V=23$ Mpc. One can see the onset of pathological behaviour for the cubic galileon case with $\delta_v = -0.7$ (top panel, red curve). All void profile parameters except $\delta_v$ are held at the default values of eq.(\ref{complex_params}).}
\label{fig:dvvar}
\end{center}
\end{figure*}
Fig.~\ref{fig:dvvar} compares a set of tangential shear profiles for several void depths in the cubic and quartic galileon models, for the density profile of eq.(\ref{complex_profile}). One can see the onset of the singularity at $r/R_V\sim 0.6$ in the cubic galileon void with $\delta_v=-0.7$, whilst the corresponding quartic galileon profile remains pathology-free. We set the shear profile to zero within the pathological regime. This shift in boundary conditions substantially affects the shear profile outside the pathological regime.

\begin{figure*}
\begin{center}
\subfigure{\includegraphics[scale=0.55]{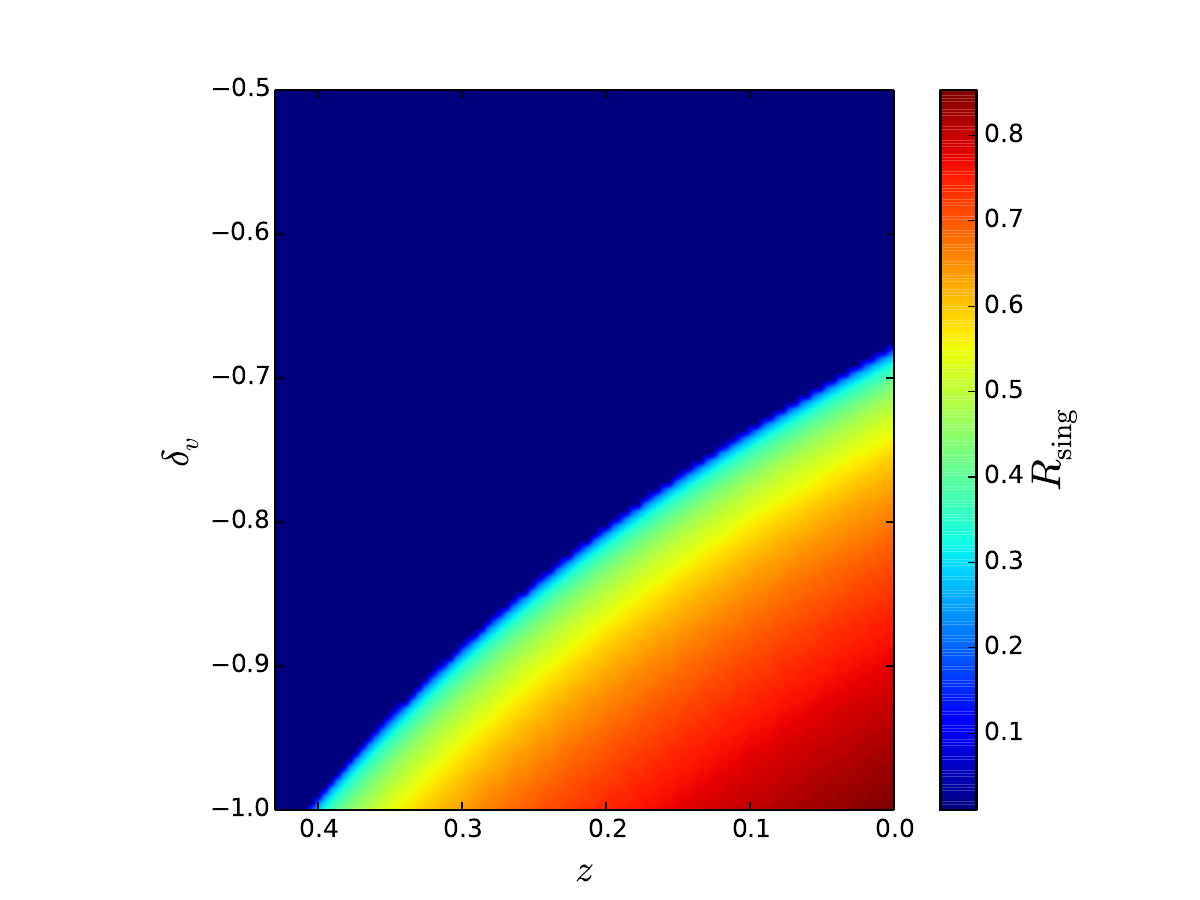}}%
\end{center}
\caption{Parameter space showing the onset of pathologies in the cubic galileon, for the density profile of eq.(\ref{complex_profile}). The colour map indicates the radius of the pathological region, in units of $R_V$. E.g., for the dark blue regions no pathologies are present anywhere throughout the void; for the reddest region in the lower right corner, pathologies are present for all radii $0<R_{\rm sing}\lesssim 0.8$, where $R_{\rm sing}=r_{\rm sing}/R_V$.}
\label{fig:grid}
\end{figure*}

Fig.~\ref{fig:grid} shows the onset of pathological behaviour for the cubic galileon, in a parameter space of void redshift and central density. For $\delta_v\leq -0.7$, the regime of complex solutions first begins at small radii inside the void, and expands outwards as $z\to 0$. For deeper voids, the onset of the pathologies occurs at higher redshifts.

The colourscale of Fig.~\ref{fig:grid} indicates the smallest radii for which real solutions for $\phi_{,\chi}/\chi$ exist; we denote this in units of the void radius as $R_{\rm sing} = r_{\rm sing}/R_V.$ This plot is constructed using the flexible void density profile, eq.(\ref{complex_profile}), with the default parameter values of eq.(\ref{complex_params}). However, we find that the onset of the pathological regime has only a very weak dependence on the void density profile, at least for the two profiles used in this paper.

The significance of this pathological regime is a particularly pressing question. One possibility is that it signals a breakdown of one of the assumptions used in the calculation of \S\ref{section:voidcalc}. Alternatively, it may be a true sickness of the cubic galileon theory itself. As we explain below, it seems likely that this behaviour is related to a breakdown of the quasistatic approximation (\S\ref{subsec:quasistatic}) at late times and in deep voids. 

It was shown in~\cite{Renk2017} that, using the cubic galileon, and using the same tracker ansatz as we have used here, the gravitational potentials $\Phi$ and $\Psi$ grow at late times in the universe on cosmological scales. This is in sharp contrast to the usual behaviour of $\Lambda$CDM, in which gravitational potential wells decay after the matter era. This effect is sufficiently strong to put the cubic galileon into $\sim7.8\sigma$ tension with measurements of the galaxy-ISW cross-correlation from \textit{Planck} and the WISE survey. Fig.~1 of \cite{Renk2017} shows that the lensing potential begins to diverge away from its $\Lambda$CDM evolution below $z\simeq 0.5$, which approximately coincides with the onset of pathologies for the deepest voids, as shown in our Fig.~\ref{fig:grid}. Such rapid evolution of the gravitational potentials may violate the quasistatic assumption that their time derivatives are negligible compared to their spatial derivatives.

Indeed, one can see from the full, unapproximated set of equations presented in~\cite{Winther:2015pta} that, at late times, the temporal and spatial derivatives of $\Phi$ enter at the same order. It is easy to believe that non-negligible values of $\dot\Phi$ and $\dot\Psi$ could then completely alter the character and solution of the galileon field equations, perhaps removing the complex solutions altogether.  A concrete proof of this hypothesis requires a dedicated numerical investigation using the full equations of~\cite{Winther:2015pta}. Such work is beyond the scope of the present paper. However, our work here usefully delineates the approximate boundaries in void depth and redshift up to which the quasistatic calculations  can be used.

\section{Conclusions}
\label{section:discussion}

Cosmological tests of gravity are evolving rapidly. When new tools or observations arise, they allow us to probe current models in hitherto-unexplored regimes, sometimes uncovering stark predictions and invalidating the models in question. This information then sculpts the next suite of ideas regarding extensions of GR and mechanisms for cosmic acceleration.

Recently, gravitational waves have provided a prime example of this. The gravitational lensing of voids is an equally novel field that has similar potential for probing the behaviour of gravity in low-density environments. This regime is notably orthogonal to all established high-precision tests of gravity, since no comparable low-density environment exists inside the Galaxy. Furthermore, there is reason to believe that any gravitational fifth forces must undergo suppression in high-density environments, leaving the door open for unscreened phenomena to manifest themselves in voids.

In this paper we have studied the lensing signatures of a family of gravity models, galileons, that invoke a single scalar field with derivative interactions. The models chosen are not intended to represent realistic models of late-time cosmic acceleration; indeed, whilst this work was in preparation, the LIGO-VIRGO Consortium and collaborating experiments announced results which eliminated the quartic galileon as a viable extension of General Relativity. However, they provide relatively simple and clean examples of modifications to GR that contain higher derivative interactions. Our work should provide a useful study of the typical phenomenology that can be produced by a non-trivial scalar field propagating on the scales of tens of megaparsecs.

Let us summarise here some of our observations on how deviations from GR can affect void shear profiles:
\begin{enumerate}[i)]
\item The effective stress-energy contribution of a scalar field to the Poisson equation can produce significant deviations from GR void lensing profiles, even if  void shapes in modified gravity are very little changed from those of GR voids {\cite{Barreira2015,Cai2014,Cai2015}}. 

Variations amongst the gravity models studied here impacted the amplitude of the shear profile, but did not shift its zero-crossing; the lensing signal beyond the void radius was also typically boosted. Our finding are consistent with those of \cite{Barreira2013b}.
\item  As shown in Fig.~\ref{fig:density}, the inferred distribution of the central underdensity, $\delta_v$, is changed by approximately $20\%$ in cubic galileon gravity, compared to in GR. If one assumes that the galaxy-mass connection is unaltered, then the galaxy tracer profile of voids can be used to estimate $\delta_v$ independently, thus providing a consistency test of the theory. In future work, reliable mock catalogs and  dedicated, model-specific simulations of alternative gravity theories will be needed to obtain high-confidence constraints. 
\item The widely-used quasistatic and tracker approximations have significant impacts on void lensing. The tracker approximation implies theoretical constraints (\S\ref{subsec:tracker}) which can affect the lensing amplitude in a non-intuitive way, by fixing some of the gravity model parameters. The quasistatic approximation is the likely cause of the mathematical singularities occurring in the deepest voids at low redshift. One should examine the behaviour of \textit{any} gravity theory carefully in this regime (low $z$ and large $|\delta_v|$) before applying a quasistatic treatment.
\end{enumerate}
Of course, it is possible that gravitational models other than those studied here could produce different phenomenology. On large cosmological scales, where linear perturbation theory is valid, one may build unified frameworks that encompass many gravity different models, and use these to perform generalised calculations \cite{Baker2013,Gleyzes2014,Gleyzes2016,Lagos2016,Lagos2017a,Lagos2017b}. At present there is not an equivalent treatment applicable to fully non-linear scales (although for a recent idea see \cite{Cusin2017a,Cusin2017b}); the best one can do is to investigate an array of models, and attempt to extract general common behaviours. 

From an observational viewpoint, the overall enhancement of void lensing -- plus the boost in the shear profile outside the void radius for compensated voids -- provide a potentially accessible handle on these theories of gravity. The errors on the tangential shear measured by the full Dark Energy Survey (DES) \cite{Flaugher2005,DES2005} will be at least a factor of two smaller than those published in the DES Year 1 analysis, which indicates encouraging prospects for constraining the phenomenology seen here. Likewise, the abundance of voids itself provides a route to constrain the dark energy parameters $\{w_0,w_a\}$; forecasts for the Euclid and WFIRST \cite{Spergel2015} satellites are given in \cite{Pisani2015}. 

Void lensing with the Euclid and LSST surveys will provide major advances in the statistical errors and redshift coverage of the void sample. With voids that extend to $z \sim 1$, one may also pursue possible evolution in the lensing enhancement due to modified gravity theories. We hope to forecast the potential of voids in these surveys to constrain deviations from GR in a future work.

\section*{Acknowledgments}
\noindent It is a pleasure to thank Alex Barreira for extensive discussions and code comparison efforts during this work. We would also like to thank Chen Su for spotting an error in an earlier version of this manuscript. BJ is grateful to Arka Banerjee, Eric Baxter, Daniel Gruen, Yuedong Fang, Oliver Friedrich, Elisabeth Krause, Nico Hamaus, Andras Kovacs, Giorgia Pollina, Carles Sanchez and DES colleagues for discussions and related collaborative work. TB is supported by All Souls College Oxford, the US-UK Fulbright Commission and the Beecroft Trust. TB would like to thank the Center for Particle Cosmology at the University of Pennsylvania, where this work was begun, for their kind hospitality. MT would like to thank the University of Oxford for hospitality. BJ and JC are partially supported by Department of Energy grant DE-SC0007901. MT is supported in part by US Department of Energy (HEP) Award dE-SC0013528, and by NASA ATP grant NNX11AI95G.


\section*{Appendix}
\label{appendix}
\noindent Here we give expressions for the coefficients appearing in eqs.(\ref{eq:poisson}) - (\ref{eom_phi}) \cite{Barreira2013b}. All are functions of the background, homogeneous value of the scalar field, and primes indicate derivatives with respect to $\ln a$. $\xi$ is the constant parameter appearing in the tracker ansatz, eq.(\ref{eq:tracker}), and $c_3$ and $c_4$ are parameters of the galileon Lagrangian, eq.(\ref{eq:action}). Note that $\xi$, $c_3$ and $c_4$ are not all independent -- see \S\ref{subsec:tracker}.
\begin{align}
A_1 &= -2c_3\xi{\bar\phi}' -12c_4\xi^2{\bar\phi}' \\
A_2 &= 6c_4\xi{\bar\phi}' \\
A_4 &= 2 - 3c_4\xi^2{\bar\phi}'^2 
\end{align}
 
\begin{align}
B_0 &= -2 - c_4\xi^2{\bar\phi}'^2 \\
B_1 &= 4c_4\left(-\xi^2{\bar\phi}' - \frac{3}{2}\xi^2{\bar\phi}''\right) \\
B_2 &= 2c_4\xi{\bar\phi}' \\
B_3 &= -2 + 3c_4\xi^2{\bar\phi}'^2 
\end{align}

\begin{align}
C_1 &= 1 - 2c_3\left(4\xi + \xi\frac{{\bar\phi}''}{{\bar\phi}'}\right) -c_4\left(26\xi^2 + 6\xi^2\frac{{\bar\phi}''}{{\bar\phi}'}\right) \\
C_2 &= 4c_3 + 6c_4\left(2\xi + \xi\frac{{\bar\phi}''}{{\bar\phi}'}\right)  \\
C_3 &= -4c_4  \\
C_4 &= 2c_4\left(3\xi^2{\bar\phi}'' + 2\xi^2{\bar\phi}'\right) \\
C_5 &= -2c_3\xi{\bar\phi}' - 12c_4\xi^2{\bar\phi}'  \\
C_6 &= -4c_4\xi{\bar\phi}' \\
C_7 &= 12c_4\xi{\bar\phi}' 
\end{align}

\begin{align}
\eta_{01}&=-\frac{{A_4} \Omega_M}{a^3}  ({B_0} {C_5}+{B_3} {C_4})\\%
\eta_{10}&=-A_4 (A_4 B_3 C_1 + A_1 B_3 C_4 + A_1 B_0 C_5 + A_4 B_1 C_5)\\%
\eta_{11}&=-\frac{\Omega_M}{a^3}  {A_4} ({B_0} {C_7}+{B_3} {C_6})\\
\eta_{20}&=-{A_4} \left[{A_1} {B_0} {C_7}+{A_1} {B_3} {C_6}+{A_2} {B_0} {C_5}+{A_2} {B_3} {C_4}\right]\nonumber\\
&-{A_4}^2 ({B_1} {C_7}+{B_2} {C_5}+{B_3} {C_2})\\
\eta_{30}&=-{A_4} [{A_2} {B_0} {C_7}+{A_2} {B_3} {C_6}+A_4({B_2} {C_7}+{B_3} {C_3})] 
\end{align}

\bibstyle{apsrev}

\end{document}